\documentclass[aps,twocolumn,showpacs,superscriptaddress,floatfix]{revtex4}
\usepackage{graphicx}
\usepackage{psfrag}
\begin{document}

\title{The Topological Non-connectivity Threshold 
and magnetic phase transitions
in classical anisotropic long-range interacting spin systems}

\author{R.~Trasarti-Battistoni}
\affiliation{Dipartimento di
Matematica e Fisica, Universit\`a Cattolica, via Musei 41, 25121
Brescia, Italy}
\author{F.~Borgonovi}
\affiliation{Dipartimento di
Matematica e Fisica, Universit\`a Cattolica, via Musei 41, 25121
Brescia, Italy}
\affiliation{ I.N.F.N., Sezione di Pavia, Italy}
\author{G.~L.~Celardo}
\affiliation{Dipartimento di
Matematica e Fisica, Universit\`a Cattolica, via Musei 41, 25121
Brescia, Italy}

\begin{abstract}

We analyze from the dynamical point of view the classical characteristics of 
the Topological Non-connectivity Threshold (TNT),
recently introduced in 
F.Borgonovi, G.L.Celardo, M.Maianti, E.Pedersoli, 
J. Stat. Phys., 116, 516 (2004).
This shows interesting connections among
Topology, Dynamics, and Thermo-Statistics of 
ferro/paramagnetic phase transition in classical spin systems,
due to the combined effect of anisotropy and long-range interections.

\end{abstract}
\date{\today}
\maketitle

\section{Introduction}

The magnetic properties of materials are usually described in the frame
of system models, such as Heisenberg or Ising models where rigorous results,
or suitable mean field approximations are available in the 
thermodynamical limit.
On the other side, modern applications require to deal with  nano-sized
magnetic  materials, whose intrinsic features lead, from one side
to the emergence of quantum phenomena\cite{chud}, and to the other
to the question of applicability of statistical mechanics.
Indeed, few particle systems do not usually fit in the class of systems
where the powerful tools  of statistical mechanics can be applied 
at glance. In particular, an exhaustive theory able to fill the gap
between the description of  $2$ and $10^{23}$  interacting particles
is still missing.
Moreover, also important well-established thermodynamical concepts
as the temperature, become questionable at the nano-scale\cite{tempe}.
In a similar way, long-range interacting systems belong,
since long, to the class where standard statistical mechanics cannot
be applied {\it tout court}. 
Indeed, they  display a number
of bizarre behaviors, to quote but a few, 
negative specific heat \cite{Lynden-Bell} 
and hence ensemble inequivalence \cite{InEq},
temperature jumps, 
and long-time relaxation (quasi-stationary states)\cite{QSS}.
Therefore, from this  point of view, few-body short-range interacting
systems share some similarities with many-body long-range ones.

Within such a scenario, and thanks to the modern computer capabilities,
it is quite natural take a different point of view,  starting
investigations directly from the dynamics, 
classical and quantum as well \cite{celardo}.
It was in this spirit that quite recently 
in a class of anisotropic Heisenberg-like spin lattice systems,
a topological non-connection of the phase space was discovered \cite{jsp}. 
Initially, for historical reasons \cite{palmer}, 
this was referred to as broken ergodicity,
since if the phase space is decomposable in two unconnected parts
(hence topologically non-connected)
then a breaking of the ergodicity is indeed a trivial consequence 
\cite{khinchin}.
Here we prefer to call Topological Non-connectivity Threshold (TNT)
the value $E_{tnt}$ where such a disconnection sets in 
upon lowering the total energy $E$ of the system.
This result was found, first numerically and later analytically,
in a class of spin models where important and rigorous results 
have been obtained during the last century, 
though generally only in the thermodynamical limit
(total number of microscopic constituents of the system $N\to\infty$
keeping fixed the number density $N/R^d$, 
where $R$ is the spatial size of the system and $d$ 
is its spatial embedding dimension).
Nevertheless, to the best of our knowledge, 
apparently nobody took care of the dynamics, 
and consequently nobody spoke of this simple but relevant property.

This dynamical point of view has a few interesting classical consequences.
First of all it explains, from the  point of view of microscopic dynamics,
the possibility of ferromagnetic behavior in small system.
Indeed,  in absence of external field and
external noise (temperature) a magnetized  system, (belonging
to one branch of the non-connected phase space)
remains magnetized simply because it cannot move 
to the other one. 
Furthermore, our TNT is surely related to 
recent results\cite{pettini,HK,ARZ} connecting 
topological transitions (TT) and thermo-statistical phase transitions (PT),
even if such investigations again concern the thermodynamical limit only,
and they relate to usual PT of canonical thermo-statistics.
However, 
it has been recently stressed \cite{Gross}
that microcanonical thermo-statistics is the theoretically more suitable
description for systems with small size and/or long-range interactions.

We therefore begin in Section \ref{mechtop} 
with a short description of our class of models 
and the topological properties of the TNT for finite and infinite $N$,
pointing out the crucial role played by the first
key ingredient of our models, the $XY$ anisotropy 
In particular we would like to answer the following question: 
what happen to the energy $E_{tnt}$
and to the corresponding $N$-spin configuration $\vec S_{tnt}^N$
when the $N\to\infty$?
It is exactly answering this question that the deep connection
between TNT and the other key ingredient, 
the long-range interactions, becomes apparent.
Our results \cite{brescia} can be summarized as follows:

1) For finite sized systems TNT exists for short range
and long range as well, as soon as anisotropy persists.

2) When the number of particles becomes large
the energy size  of the non-connected  region
goes to zero with respect to the total energy size for short
range interaction while it goes to some finite quantity
for long range ones.

In Section \ref{dynthermostat} 
we switch to the dynamical properties found in \cite{firenze} 
which accompany such a special (and rather ``big'') TT,
and its relations with its thermostatistical properties,
namely the occurence and location of a usual (paramagnetic/ferromagnetic) PT,
using techniques from large deviation theory \cite{dauxois} 
within the microcanonical description of the system.

We conclude in \ref{conclusions}.
Everywhere here we restrict ourselves to the classical case;
for a recent discussion of its quantum counterpart,
we refer to \cite{bcb,bct}.

\section{Mechanics and Topology}
\label{mechtop}

As a paradigmatic model example of TNT,
let us consider the following class of lattice spin models, 
described by the Heisenberg-like Hamiltonian:

\begin{equation}
\label{H}
H=
  {\eta \over 2}\sum_{i\not=j}^N {S^x_i S^x_j \over r_{ij}^\alpha}
-      {1 \over 2}\sum_{i\not=j}^N {S^y_i S^y_j \over r_{ij}^\alpha}
\end{equation}

\noindent
where  
$S^x_i, S^y_i, S^z_i$ are the spin components, assumed to vary continuously; 
$i,j=1,...N$ label the spins positions 
on a suitable lattice of spatial dimension $d$,
and $r_{ij}$ is the inter-spin spatial separation therein.
Here for simplicity we consider a $d=1$ lattice.
(See \cite{brescia} for extensions to $d=2,3$.)
For later notational convenience, 
we define the single-spin vector $\vec S=(S^x_i, S^y_i, S^z_i)$ 
and the $N$-spin configuration $\vec S^N=(\vec S_1,\vec S_2,...\vec S_N)$.
The tip of each $i$-th spin is taken to lie on the unit sphere, i.e. 
it satisfies the constraint $(S^x_i)^2 + (S^y_i)^2 +  (S^z_i)^2 = 1 $.
Also, 
$\alpha  > 0 $ parametrizes the range of interactions
(decreasing range for increasing $\alpha$)
and 
$0< \eta< 1 $ parametrizes the $XY$ anisotropy.
For $\alpha=\infty$ we recover nearest-neighbor interactions,
while $\alpha=0$ corresponds to infinite-range interactions.
A mean-field model is obtained  by setting $\alpha=0$ and including 
as well the (somewhat non-physical) self-interaction pairs $i=j$:

\begin{equation}
\label{hmf}
H_{mf}  = \frac{\eta}{2} N^2 (m_x)^2 - \frac{1}{2} N^2 (m_y)^2,
\end{equation}
where 
$
m_{x,y,z} = (1/N)\sum_i^N S^{x,y,z}_i.
$
While this might be thought of as a negligible modification for $N\to\infty$,
nevertheless it has non-negligible effects concerning
the chaoticity properties of the system.
Indeed,
the dynamics of the truly mean-field $\alpha=0$ system
turns out to be exactly integrable\cite{firenze}.
Here we are not interested in the most general spin Hamiltonian
giving rise to a TNT (for instance in \cite{jsp, firenze} 
a term containing a transversal magnetic field $B_z$ has been added to $H$). 
Rather, 
we focus on the very simple Hamiltonian (\ref{H}) which already contains 
the two essential ingredients which give rise to the TNT, 
i.e. anisotropy and long-range,
whose main effects are conveniently encapsulated within two
simple and easy-to-handle parameters $\alpha$ and $\eta$,
in order to make the study of the minimum-energy configurations
analytically tractable at $N\to\infty$
and numerically feasible at finite $N$.
Concerning the physical motivations for such choices,
we refer to the recent discussion in \cite{bct}.

Depending on the specific $N$-spin configuration 
the corresponding energy $E=H(\vec S^N)$ will vary according to (\ref{H}).
One (not necessarily unique) configuration $\vec S^N_{min}$, 
to be specified soon, turns out to have minimum energy $E_{min}$, 
defined by the minimum of the Hamiltonian (\ref{H}) 
over all conceivable configurations:

\begin{equation}    
\label{emin}
E_{min} = Min \ [ \ H \ ]\ .
\end{equation}

\noindent
Since $0 < \eta < 1$ the minimum energy configuration $\vec S^N_{min}$
is attained when all spins are aligned along the $Y$ axis, which
defines implicitly the easy axis of magnetization.
In the same way, let us  define the TNT energy $E_{tnt}$ as 
the minimum energy compatible with the constraint of 
zero magnetization along the easy axis of magnetization:

\begin{equation}
\label{etnt}
E_{tnt} = Min \ \left[ \ H  \ \left| \ m_y  =  0 \ \right]   \right..
\end{equation}

\noindent
corresponding to some other $N$-spin configuration $\vec S^N_{tnt}$ 
to be specified later.
By definition, in general $E_{min} \le E_{tnt}$,
and in particular it may be that $E_{min} \ne E_{tnt}$. 
We call this situation is topological non-connection,
as will become clear in a moment,
and its physical (dynamical as well as statistical) consequences 
are rather interesting. 
Indeed, 
consider a system prepared at time $t=0$
with a definite sign of magnetization, say $m_y>0$ 
and an energy value $E_{min} \le  E \le   E_{tnt}$.
As time goes by,
the system evolves upon the constant energy surface $H(\vec S^N)=E$ 
in configuration space.
Nevertheless,
due to the continuity of the dynamical equations of motion 
the magnetization $m_y(t)$ (not a constant of motion) 
may well change its size, but it can never change its sign, instead.
Indeed, 
in order to assume a value $m_y<0$ 
it should have to go through at least one configuration with $m_y=0$,
which by definition cannot belong to the $E<E_{tnt}$ surface.
The whole situation can be summarized as follows.

{\it Topology}: 
in configuration space the surface at fixed energy $E$
is topologically non-connected in two components, 
each characterized by a magnetization either $m_y<0 $ or $m_y>0$.

{\it Dynamics}: 
though the two components are energetically accessible on equal grounds,
the ergodicity of the constant $E$ surface is trivially broken,
since there exist no dynamically allowed path inbetween them.

{\it Thermo-Statistics}: 
de-magnetization is in principle impossible below the TNT,
so  we may speak in some sense of a ferromagnetic phase.
Of course, the application of a magnetic field, or a thermal noise,
can give the energy necessary to overcome the energy barrier,
thus in principle - if not in practice - allowing for a magnetic reversal.
On the contrary, for energy values $E>E_{tnt}$,  
de-magnetization is in principle possible,
and we may speak in some sense of a paramagnetic phase.
However, 
being above the TNT does not automatically guarantee that, 
for any combination of parameter values and initial conditions, 
a system initially magnetized at $t=0$ 
will for sure eventually de-magnetize 
within a given finite observational time $t_{obs}$.
First, 
as reported in \cite{firenze} 
some invariant multidimensional structures can appear in phase-space
preventing the motion from covering the whole energy surface.  
This lack of ergodicity is well known in 2D  Hamiltonian systems, 
where KAM tori prevent motion from wandering the whole phase space
\cite{chirikov}, 
while it turns out to be more complicated in multidimensional system, 
see for instance the famous Fermi-Pasta-Ulam problem
(for interesting reviews see \cite{ford,felix}).
Such invariant structures usually disappear 
in the presence of strongly chaotic motions 
\cite{chirikov}. 
We can therefore say that strong chaos 
is somehow another necessary ingredient 
in order for the system to be in its paramagnetic phase.
Second,
even given strong enough chaoticity to ``encourage'' the system
to explore all the available phase (or configuration) space,
yet the system could be given not enough time to actually do it,
so effectively ``freezing'' it within the $m_y$ component 
where it started from.

For finite $N$ systems the $XY$ anisotropy 
is the only necessary ingredient in order to have $E_{tnt}<E_{min}$
and hence the TNT.
For $N \to \infty$, 
one quickly realizes that $E_{min}\to-\infty$
and guesses that $E_{tnt}\to-\infty$,
but may still wonder wehther $E_{tnt}\to E_{min}$ as well,
thus making the TNT physically irrelevant.
So we define the topological non-connection ratio:

\begin{equation}
\label{erre}
r = \frac{ E_{tnt} - E_{min}}{|E_{min}|},
\end{equation}

\noindent
which expresses how large a fraction of the energy range
is topologically non-connected.
Here we use the denominator $E_{min}$ 
(instead of the whole energy range), since, in such kind of models,  
one is usually interested in the negative energy range only.
Correspondingly,
we will refer to a system as topologically non-connected
if $r \to const \ne \ 0$ in the limit $N \to \infty$. 

To show why and how this whole comes by in our models,
we now concentrate on the energy $E$ of the above mentioned 
interesting $N$-spin configurations.
The minimum-energy configuration $\vec S^N_{min}$
(all spins aligned along the $Y$-axis),  can also be thought of 
as composed of $2$ equal blocks of $N/2$ spins, all up.
Correspondingly, 
the minimum energy $E_{min}<0$ can be split into 3 contributions,
namely $2$ equal intra-block bulk energies $E_\uparrow<0$
plus $1$ inter-block interaction energy $V_{\uparrow\uparrow}<0$.
Flipping just 1 block, 
its bulk energy does not change, i.e. $E_\uparrow=E_\downarrow<0$,
but the interaction energy reverses sign to
$V_{\uparrow\downarrow}=-V_{\uparrow\uparrow}>0$.
However, 
in such a new configuration $\vec S^N_{\uparrow\downarrow}$
the magnetization is now $m_y=0$.
Both bulk energies are as low as they can,
while the constraint $m_y=0$ frustrates the total energy 
to rise above $E_{min}$ by an amount $2|V_{\uparrow\uparrow}|$.
We therefore take  $E_{\uparrow\downarrow}=E_{min}+2|V_{\uparrow\uparrow}|$
as a  somewhat obvious candidate for $E_{tnt}$ and the 
related spin configuration $\vec S^N_{\uparrow\downarrow}$ as the 
TNT spin  configuration $\vec S^N_{tnt}$.

\begin{figure}
\includegraphics[scale=0.40]{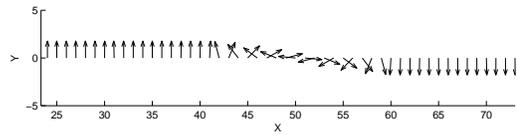}
\caption{ 
TNT spin configuration. Here is $\alpha=0.1$, $N=100$,
$\eta=0.9$. Only the central part of the chain has been shown.
}
\label{n100}
\end{figure}

\noindent
A $d=1$ example of a numerically obtained $\vec S^N_{\uparrow\downarrow}$ 
is given in Fig.~\ref{n100}.
As one can see, excluding a small central domain, 
the spins are arranged in two equal-size blocks with opposite spins,
i.e.
the real $\vec S^N_{tnt}$ is remarkably similar to,
though in fact somewhat slightly different from,
the ideal $\vec S^N_{\uparrow\downarrow}$.
Extensive numerical simulations using constrained optimization 
and analytical estimates as well\cite{brescia}, 
confirm such an impression, with natural analogues in $d=2,3$.

Quite generally,
the exact TNT configuration at finite $N$
for arbitrary $\alpha$ and $\eta$ can be found only numerically.
Nevertheless, 
concerning the large $N$ limit it is possible
to make a few definite statements \cite{brescia}.
First, 
if $\alpha \ne 0$, $\eta=\-1$, and  $N$ sufficiently large, 
then 
$\vec S^N_{\uparrow\downarrow} \simeq \vec S^N_{\uparrow\downarrow}$,
in the twofold sense that 
the TNT spin configuration is approximately half up and half down,
and the energy difference 
$E_{dom}:=E_{tnt}-E_{\uparrow\downarrow}$ due to the domain
is a finite quantity independent of $N$. 
Second, for $N\to\infty$,
if $\alpha>d$ (short-range) then $r\to 0$
but
if $\alpha<d$ (long-range) then $r\to const \ne 0$.
In particular, if $d=1$ then $r_\infty=2-2^\alpha$.
Intuitively,
while for short-range the bulk volume contribution
$E_\uparrow + E_\downarrow$ 
eventually dominates over the geometric inte(sur)face area
energy interaction $V_{\uparrow\uparrow}0$,
for long-range interactions the ``effective interface'' 
is the whole bulk as well.
As now both
the inter-block and the intra-block interaction 
involve all spin pairs in the two blocks,
though each pair with a different intensity $\propto r_{ij}^{-\alpha}$,
energies will then scale like 
the number of pairs $\propto N^2$ times 
the interaction at typical distance $r_{ij}\simeq R$.
So there holds the {\it same scaling}
$|V_{\uparrow\uparrow}|\sim (R^d)^2/R^\alpha\sim N^{2-\alpha/d}\sim |E_{min}|$,
though with different (and $\alpha$-dependent) proportionality constants,
and so $r \simeq const \not \to 0$ as $N \to \infty$.

\section{Dynamics and Thermo-Statistics}
\label{dynthermostat}

Here, 
we show how to impose a self-consistent Hamiltonian dynamical evolution
upon the spin systems described by (\ref{H}).
Afterward, we will employ such dynamical equations,
numerically integrated for long enough times,
to study the time-evolution of $\vec S^N(t)$.
Following \cite{brescia, firenze}
here we focus on the time evolution instantaneous magnetization $m_y(t)$
started with some $E$ and $m_y(0)$ and evolved according to (\ref{eqm}).
Complementarily,
we look at its statistical distribution $P_{ens}(E,m_y)$,
built via a random sampling of an ensemble of initial conditions, 
all with the same $E$, tunable at will.
Finally,
we will show a connection between Dynamics and Thermo-Statistics.

For each $i$ the spin components $S^x_i,S^y_i,S^z_i$
satisfy the usual commutation rules for angular momenta,
i.e. 
$\{S^x_i,S^y_j\}=\delta_{ij} S^z_i$, (and cyclic)
 where $\{,\}$ are the canonical Poisson brackets.
As usual,
(and this procedure immediately translates to the quantum case \cite{bcb,bct}),
starting from the Hamiltonian (\ref{H})
we straightforwardly derive the dynamical equations as:

\begin{equation}
\label{eqm}
\nonumber \frac{d \vec S_i }{dt} = \{ H, \vec S_i \}
\end{equation}

\begin{figure}
\includegraphics[scale=0.40]{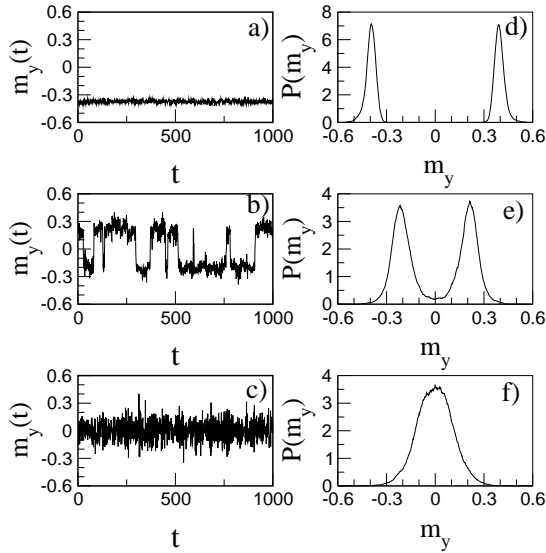}
\caption{Left column (a,b,c) : time evolution of the 
magnetization $m_y(t)$ for different specific energy values.
Right column (d,e,f) : probability distribution of the magnetization
at a given specific energy. Parameters are $\alpha=0$, $\eta=1$, $N=10$.
Upper line ((a,d) below the TNT  ) : $E/N= -0.7$.
Middle line ((b,e) between the TNT and the statistical thresholds $E/N= -0.3$. 
Lower line ((c,f) above the statistical threshold) $E/N= 0.1$.
}
\label{dyn5}
\end{figure}

\noindent
As is well-known, for such dynamical equations 
the total energy $E$ and the spin moduli $|\vec {S}_i|^2=1$ are
constants of the motion. 
Usually, for energies $E$ not too close to $E_{min}$,
the dynamics is characterized by a positive largest Lyapunov exponent,
corresponding to strongly chaotic motion \cite{jsp}.
On the contrary,
the dynamics of the truly (all pairs $ij$) mean-field $\alpha=0$ system 
turns out to be exactly integrable\cite{firenze}.
Given strong enough chaoticity,
typical examples of evolution curves of $m_y(t)$ 
and corresponding $P_{ens}(E,m_y)$, are shown in Fig.~\ref{dyn5}.

Again, the whole situation can be summarized as follows.

{\it Thermo-Statistics}: 
for $E < E_{tnt}$ (Fig.~\ref{dyn5}d) 
the two peaks of the distribution are well separated, 
while for $ E > E_{tnt}$ (Fig.~\ref{dyn5}e) they come close. 
At $E_{stat}$ they merge into one single peak (Fig.~\ref{dyn5}f),
as expected from the statistical analytical estimate.
Note that this simple (and, in its roots, purely geometrical)
argument immediately prooves that, in general, 
$E_{tnt} < E_{stat}$: 
excluding catastrophic situations, upon increasing $E$ 
the two separated inner tails must first touch at $E_{tnt}$ 
before the relatively outer twin peaks may eventually merge at $E_{stat}$.
Indeed, 
this very same argument automatically applies to 
any model with probability distribution
changing from single to double-peaked ,
e.g. the discrete mean field $\varphi^4$ model \cite{HK}.
In other words, 
if both a TNT and a PT are known to exist, 
then in general they will occur at different energies.
And depending on wether $E$ is increased or decreased,
both a TNT and a PT in some sense ``anticipate'' each other.
Of course, 
for a given specific system nothing automatically guarantees that 
either a TNT or a PT do indeed both exist.
So, while in principle a TNT and a PT neither imply nor exclude each other,
in practice they ``strongly suggest'' each other.
 
{\it Dynamics}:
at low energy (a) 
the system is always magnetized with small fluctuations, 
and at high energy (b) 
the system is on average non-magnetized with large fluctuations.
In between (c) the magnetization jumps erratically up and down.
We can usefully define a magnetic-reversal time-scale \cite{firenze}, 
as the average time necessary to magnetization to reverse its sign. 
In the presence of strong chaos (dependent on the parameters $N$ and $E$)
magnetic reversals occur fully at random, 
with the distribution of jumping times following a Poisson distribution.
Any deviation from such distribution, 
for large $N$ and large negative $E$ values, 
signals the presence of invariant curves 
preventing the motion to switch from one peak to the other one.

\begin{figure}
\centerline{\includegraphics[scale=0.35]{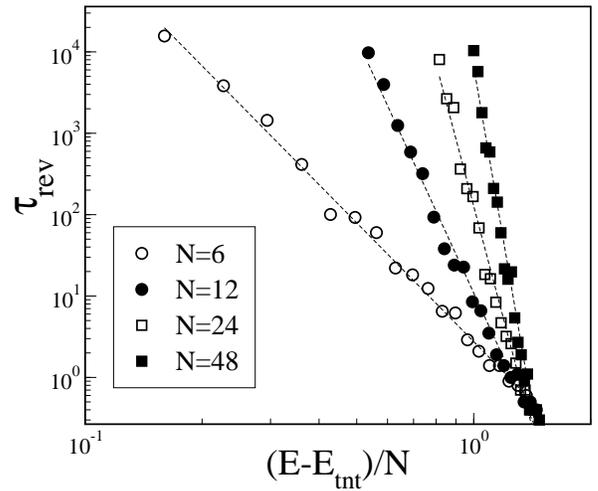}}
\caption{
Divergence of magnetization reversal times close
to the TNT. Here is $\alpha=0$, $\eta=1$, and different $N$
values as indicated in the insert. Lines are the best fit
according to Eq.~(\ref{taurev}).
}
\label{tau2}
\end{figure}

Interestingly, and somewhat reminiscent of a critical PT,
as shown in Fig.~\ref{tau2} 
the magnetic-reversal time-scale diverges 
as a power law of $E$ at the critical energy $E_{tnt}$:

\begin{equation}
\label{taurev}
\tau_{rev} \sim \left[ {E - E_{tnt} \over N} \right]^{-\gamma(N)}
\end{equation}

\noindent 
where $\gamma(N) \propto N$.

Remarkably, such a {\it dynamical} behavior can be qualitatively - 
and almost quantitatively - 
explained only by {\it statistical} considerations. 
Before that,
let us remark that (\ref{taurev}) 
can be inverted for $E$ at any fixed $N$,
thus giving a $\tau$-depending characteristic energy
$E_{rev}(\tau)$.
We will see the physical meaning of $E_{rev}$ in a while;
here we just note that, 
since (\ref{taurev}) holds only for $E>E_{tnt}$, 
it will also be $E_{rev}>E_{tnt}$.

{\it Thermo-Statistics}: 
restricting attention to the mean field Hamiltonian (\ref{hmf}), 
a detailed statistical analysis using techniques from large deviations theory
leads to definite predictions (see \cite{firenze} for details)
concerning
the microcanonical (i.e., purely state-counting at fixed $E$) 
probability distribution $P_{stat}(m_y,E)$,
the topological energy threshold $E_{tnt}$, and
the thermo-statistical energy threshold $E_{stat}$, 
where the system undergoes a continuous ferro/paramagnetic PT
in the microcanonical description.

{\it Dynamics}: 
Adopting a viewpoint as close as possible to the experimental one, 
we introduce an observational time $t_{obs}$, 
during which the experiment can be performed.
Assuming the experimentally measured value 
and the dynamically observed time-average to coincide,
we introduce the $t_{obs}$-averaged magnetization:

\begin{equation}
\label{mobs}
\langle m_y \rangle = \int_0^{t_{obs}} \ dt \ m_y(t) .
\end{equation}

\begin{figure}
\includegraphics[scale=0.35]{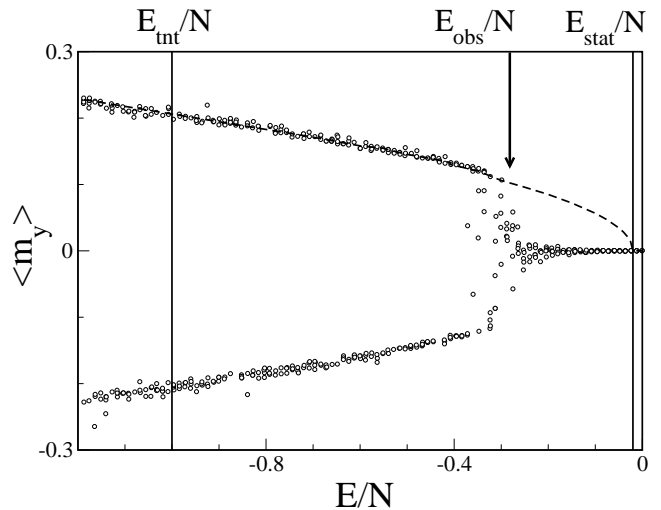}
\caption{
Numerical $t_{obs}$-averaged dynamical magnetization 
as a function of the specific energy $\epsilon=E/N$ (circles). 
Dashed line are the analytical predictions 
from the thermo-statistical mean field model (\ref{hmf}). 
Here is $N=60$, $\alpha=0$, $\eta=1$. 
For the sake of comparison, both
the thermo-statistical phase transition
and the topological non-connectivity thresholds 
have been indicated as vertical lines. 
Also indicated, as  a vertical arrow, 
the expected dynamical magnetic-reversal threshold. 
}
\label{stat}
\end{figure}

\noindent
The quantity $\langle m_y \rangle$ as a function of the energy $E$ 
can be compared with $\hat m_y(E)$, 
obtained from 
the most likely values $P_{stat}(E,\hat m_y)$ 
(the maxima of the distribution).
This is shown in Fig.~\ref{stat}, where we have also indicated
the thresholds $E_{stat}$ and $E_{tnt}$ as vertical lines.

Of course, below $E_{tnt}$ 
and above $E_{stat}$ the two averages 
coincides, at variance with the region between them.
Also,  
the effective  transition given by $\langle m_y \rangle$
occurs at some intermediate energy, 
different from both $E_{tnt}$  and $E_{stat}$.
This is not completely surprising, 
since both the topological energy $E_{tnt}$ 
and the thermo-statistical energy $E_{stat}$ 
refer to some limiting properties, 
the former when $t \to \infty$ at any finite $N$,
the latter when $N \to \infty$ at any time $t$.
It is interesting to note that, 
inverting  Eq.~(\ref{taurev}),
to the observation time $t_{obs}$ 
there correspond a dynamical energy $E_{rev}(t_{obs})$ 
which is in  fairly good agreement
with the observed transition energy, see Fig.~\ref{stat}.

Note also that, to a rather good accuracy \cite{firenze,bcb},
there holds the following proportionality
between the {\it dynamical} quantity $\tau_{rev}$
and the {\it statistical} quantity $P(E,m_y)$:

\begin{equation}
\label{tauPP}
\tau_{rev} \propto {\hat P \over P_0}
\end{equation}

\noindent 
where $\hat P=P_{stat}(E,\hat m_y)$ and  $P_0=P_{stat}(E,0)$.
Theoretically \cite{firenze},
this can be traced back to the occurence of an entropy barrier 
$\Delta S :=\hat S - S_0$ between the two components,
where $\hat S= \ln \hat P$ and $S_0= \ln P_0$,
under the assumption of markovicity of the dynamical evolution.
Detailed statistical consideration 
on the mean field model (\ref{hmf})
lead to the analytical estimate $\gamma_{stat}(N) =N$ 
to be compared with
the numerical result $\gamma_{stat}(N) \approx  0.88 N$.

Heuristically, (\ref{tauPP}) can be understood as follows.
Suppose the dynamics is strongly enough chaotic,
making the system almost freely wander through configuration space,
in some sort of diffusive fashion.
Then,
the probability to encounter a configuration with $m_y=0$,
and therefore presumably 
switch from the $m_y>0$ to the the $m_y<0$ component,
can be assumed to be simply proportional to the ratio
of the number $M_0$ of ``right'' configurations at $m_y=0$
to the number $M_+$ of ``wrong'' configurations at $m_y\ne 0$,
but counting only a half of the total wrong ones, 
namely those on the side where the system is coming from.
(The system cannot know nothing yet about the other side still it's going to.)
However,
as long as the magnetization peaks are high and narrow,
essentially all the ``wrong'' configuration are actually
within the peak itself, i.e. normalizing by the total number
$M=M_+ + M_0+ M_-$ of energetically allowed configurations
gives the probabilities $P_+ \simeq \hat P$.
Finally,
the average time needed for magnetic reversal
can be assumed to be taken proportional to
the probability to encounter the right configuration with $m_y=0$,
times some characteristic ``residence time'' $\tau_{res}$
during which the system essentially stays within each given 
$N$-spin configuration, 
along its longer wandering of the whole $N$-spin configuration space.
Altogether, 
such a purely statistical argument recovers exactly (\ref{tauPP}),
implicitly suggesting that the precise value of $\tau_{res}$
might be the origin of the different coefficients
found between the classical \cite{firenze} and quantum \cite{bcb}
realizations of the same dynamical evolution equations (\ref{eqm}).

\section{Conclusions}
\label{conclusions}

We showed the existence in classical Heisenberg spin models of
a Topological Non-connectivity Threshold (TNT), caused  by the
anisotropy of the coupling when it induces 
an easy-axis of the magnetization.
Below the TNT the constant energy surface is 
topologically disconnected in two symmetrically equal components.

Such a result on the Topology has connections both with
the Dynamics (time-scales)
and with the Thermo-Statistics (PT) of the system.
In each component the magnetization along the easy axis
has a definite sign, and it cannot change sign as time goes by,
corresponding a ferromagnetic behavior of the system.
Above the TNT, in a strong chaotic regime, 
the magnetization reverses its sign with a characteristic time-scale
which diverges as a power law at the TNT.
Given enough chaos and enough observational time,
this leads to a paramagnetic behavior of the system.
Moreover, 
the numerically observed link between 
time-scales and probability distributions 
has been given an heuristical justification.

The connection between the TNT and the range of the 
interaction has also been shown.
For macroscopic spin systems
the existence of this threshold determines 
a disconnected energy range that remains relevant
(w.r.t. the total energy range) for long-range interactions,
while it goes to zero for short-range interactions.


\begin{thebibliography}{}

\bibitem{chud} E.~M.~Chudnovsky and J.~Tejada, {\it Macroscopic Quantum
Tunneling of the Magnetic Moment}, Cambridge University Press,
(1998).

\bibitem{tempe} M.~Hartmann, G.~Mahler and O.~Wess. 
Phys. Rev. Lett. {\bf 93}, 80402 (2004).

\bibitem{Lynden-Bell}
D.Lynden-Bell, R.Wood, Mon. Not. R. Astr. Soc. {\bf 136}, 101 (1967);
W.Thirring, Z. Phys. {\bf 235}, 339 (1970);
D.Lynden-Bell, R.M.Lynden-Bell, Mon. Not. R. Astr. Soc. {\bf 181}, 405 (1977);
D.Lynden-Bell, cond-mat/9812172.

\bibitem{InEq} J.~Barr\'e, D.~Mukamel, S.~Ruffo, Phys. Rev. Lett. {\bf 87}, 3, (2001).

\bibitem{QSS} J.~Barr\'e, F.~Bouchet, T.~Dauxois, and S.~Ruffo, Phys. Rev. Lett. {\bf 89}, 110601, (2002)

\bibitem{palmer} R.~G.~Palmer, Adv. in Phys., {\bf 31}, 669 (1982).

\bibitem{khinchin} A.~I.~Khinchin {\it  Mathematical Foundations of
Statistical Mechanics}, Dover Publications, New York (1949).

\bibitem{pettini}
L.Caiani, L.Casetti, C.Clementi, M.Pettini, Phys. Rev. Lett. 79, 4361 (1997);
L.Casetti, M.Pettini, E.G.D.Cohen, Phys. Rept. 337, 237 (2000);
L.Casetti, M.Pettini, E.G.D.Cohen, J.Stat.Phys. 111, 1091 (2003);
R.Franzosi, M.Pettini, Phys. Rev. Lett. 92, 060601 (2004);
R.Franzosi, M.Pettini, L.Spinelli, cond-mat/05005057.
R.Franzosi, M.Pettini, cond-mat/05005058.

\bibitem{HK}
M.Kastner, Phys. Rev. Lett. 93, 150601 (2004);
I.Hahn, M.Kastner, cond-mat/0506649;
I.Hahn, cond-mat/0509136; M.Kastner, cond-mat/0509206.

\bibitem{ARZ}
L.Angelani, G.Ruocco, F.Zamponi, Phys. Rev. E {\bf 72}, 016122 (2005).

\bibitem{Gross}
D.H.E.Gross, Phys.Rept.279, 119 (1997);
D.H.E. Gross 
{\it Microcanonical Thermodynamics: Phase Transitions in Small Systems}, 
Lecture Notes in Physics {\bf 66}, World Scientific, Singapore, 2001;

\bibitem{celardo} G.~L.~Celardo, PhD dissertation, 
University of Milano, Italy (2004).

\bibitem{jsp} F.~Borgonovi, G.~L.~Celardo, M.~Maianti, E.~Pedersoli,
J. Stat. Phys., {\bf 116},  516 (2004).

\bibitem{brescia} F.~Borgonovi, G.L.~Celardo, A.~Musesti,
and R.~Trasarti-Battistoni
cond-mat/0505209.

\bibitem{firenze} G.L.Celardo, J.Barr\'e, F.Borgonovi, S. Ruffo,
cond-mat/04010119.

\bibitem{dauxois} T.~Dauxois, S.~Ruffo, E.~Arimondo,
M.~Wilkens Eds.,Lect. Notes in Phys.,  {\bf 602},  Springer (2002).

\bibitem{bcb} F.~Borgonovi, G.~L.~Celardo, and G.~P.~Berman,
cond-mat/0506233.

\bibitem{bct} F.~Borgonovi, G.~L.~Celardo, and R.~Trasarti-Battistoni
cond-mat/0510079

\bibitem{chirikov} B.~V.~Chirikov Phys. Rep., {\bf 52}, 253 (1979).

\bibitem{ford} J.~Ford, , Phys. Rep. {\bf 213}, 271 (1992).

\bibitem{felix} G.~P.~Berman and F.~M.~Izrailev, Chaos {\bf 15}, 015104 (2005).

\end{thebibliography}
\end{document}